# The Availability and Security Implications of Glue in the Domain Name System

Zheng Wang
National Institute of Standards and Technology, USA
zhengwang98@gmail.com

*Abstract*—The Domain Name System (DNS) is one of the most fundamental components of the Internet. While glue is widely used and heavily relied on in DNS operations, there is little thinking about the necessity, complexity, and venerability of such prevalent configuration. This work is the first to provide extensive and systematic analysis of DNS glue. It discusses the availability implications of glue and proposes the minimum glue records in terms of availability. It also identifies the security vulnerabilities of glue as well as the limitations of current countermeasures. Measurements show the wide occurrences of glue redundancies and glue vulnerabilities.

## I. INTRODUCTION

The Domain Name System (DNS) is one of the most fundamental components of the Internet, providing a critical link between human users and Internet locations by mapping host names to IP addresses. Besides translating human-friendly names into IP address, DNS is also used for looking up mail servers associated with a domain, discovering the server(s) providing for a particular service on the network, reversely mapping IP addresses to domain names, etc. Nearly all today's Internet applications rely on the DNS for proper function.

The ubiquity and criticality of the DNS necessitates both the availability and authenticity of responses. Name resolution malfunction or failure may impact the Internet services and applications as well as user experience. e.g., during January 2001 users found Microsoft website in outages because all the authoritative servers for the Microsoft DNS domain became inaccessible due to a simple configuration mistake [1]. The DNS root servers also have witness massive DDoS attacks in 2002 and 2007 [2] [3] when the global Internet service was more or less disrupted. Bogus DNS response may direct users to dead or unresponsive servers to find the unavailability of service, or even to malicious servers with the risk of privacy violation or sensitive information leakage. e.g., some DNS implementation vulnerabilities were exploited to inject malicious data into the cache of DNS resolvers [4] [5]. So The importance of correct and continuous functionality of DNS worths significant efforts to improve DNS resilience and security to both scenarios.

"DNS is a very complex system, even though its rules are simple and few, and even though a new DNS protocol agent can be constructed using only a few thousand lines of software code."[6] In essence, DNS is a distributed, coherent, reliable, autonomous, hierarchical database. It consists of a fast-growing name space, which is divided into a large number of zones. The administer or owner of a domain may delegate part of its name space to another domain known as a subdomain.The root zone resides at the top of the DNS hierarchy. Starting from the root zone, delegation may take place recursively at each domain, thereby an inverted DNS tree is constructed from the top down.e.g., Top Level Domains (TLDs) such as "com"and "org" are delegated at the root zone,and Second Level Domains (SLDs) such as "foo.com" and "foo.org" are delegated at its parent zone "com" and "org" respectively. Roughly speaking, name resolution also follows the DNS tree from the top down. However, DNS complexity has its effects more far-reaching than the simple tree.

Besides the parent-child relationship between any adjacent zones in the DNS tree, there are still other administrative dependencies among some seemingly irrelative domains.Such dependencies are created by locators of the sub-name space at each delegation. The delegation locator is represented as NS resource records where the subdomain and its nameservers are provided. To facilitate the locating of the delegated subdomain's nameservers, the IP addresses of such nameservers may be also presented at the delegation point. Besides the parent-child relationship between any adjacent zones in the DNS tree, there are still other administrative dependencies among some seemingly irrelative domains.Such dependencies are created by locators of the sub-name space at each delegation. The delegation locator is represented as NS resource records where the subdomain and its nameservers are provided. To facilitate the locating of the delegated subdomain's nameservers, the IP addresses of such nameservers may be also presented at the delegation point and included in the DNS response as part of referrals to the queries. These IP addresses records (in the form of A type for IPv4 or AAAA type for IPv6) are called "glue" records. The absence of such glue records either at the pertinent zone or in the pertinent response may make resolvers initiate a name resolution for the subdomain's nameservers to get their IP addresses. In the worst case, missing glue records can even cause name resolution failure due to cyclic dependency between a zone and its descendants. If the delegated nameservers fall into domains out of bailiwick of their authoritatively served domains, the nameserver resolution may involve inter-organizational dependencies. Such delegation locator resolution adds to the DNS complexity as well as raises availability and security concerns. On one hand, missing or misconfiguration of resource records which delegation locator depends on may disrupt the availability of name resolution. On the other hand, delegation locator may be targeted or exploited by attackers to mislead name resolution towards malicious nameservers.

Consistent with the principle of "descriptive rather than prescriptive" [7], DNS was specified loosely, meaning that each implementor may differently understand, interpret, thus implement DNS protocols. Provided that the looseness of the DNS specification should be seen as a strength rather than a weakness, it should still be necessary to rule about any points that are operationally believed to avoid vulnerabilities and secure availability.

Security is not among the basic design considerations of the original DNS. In the past two decades, attacks against DNS have been identified to be increasingly extensive and destructive. As the efforts to improve the security of DNS, DNS Security Extensions (DNSSEC) [8], [9] is introduced by IETF as a set of extensions to DNS which provide data origin authentication, data integrity, and authenticated denial of existence.

As a large scale infrastructure upgrade, DNSSEC is being gradually rolled out in the last few years. In June 2009, .org became the first Top-Level Domain (TLD) to sign its zone with DNSSEC. In July 2010, the signed root zone was available. There are 1298 TLDs in the root zone in total, among which 1129 TLDs are signed and publish their DS records in the root zone by May 4th, 2016. Compared with TLDs, the adoption rate of DNSSEC is extremely low at lower DNS levels. Kalafut et al. found that of the 4,947,993 SLD DNS zone data gathered, 161 zones, which is a mere 0.003% of the total, have the DNSKEY records [10]. As the protocol and administrative complexity added by DNSSEC is non-trivial, the incremental global DNSSEC deployment is expected to be a long transition. At least in part due to the limited DNSSEC deployment so far, some seemingly natural and popular practices potentially create security vulnerabilities. Moreover, if operated obliviously, the aforementioned DNS dependency complexity may even increase such vulnerabilities. On the other hand, as a cryptographic defense, DNSSEC is far from all omnipotent, and some weakness may still be exploited by a certain type of attack such as Man-in-the-Middle (MitM) attacks.

While glue records are widely used and heavily relied on in DNS operations, there is little thinking about the necessity, complexity, and venerability of such popular practice. This work is the first to provide extensive and systematic analysis of glue records.

The paper is organized as follows. The availability implications of glue are discussed in Section II. Section III presents the security implications of glue. The measurement study is illustrated in Section IV. Some related work are presented in Section V. Finally, Section VI concludes the paper.

## II. THE AVAILABILITY IMPLICATIONS OF GLUE

### A. The diverse definitions of glue

If a zone delegates part of its name space to other administrators, such delegation (or often referred to as "zone cut") is represented as "glue" in the zone. "Glue" includes any record in a zone file that is not properly part of that zone, including nameserver records of delegated subzones (NS records), address records that accompany those NS records (A, AAAA, etc), and any other stray data that might appear. Here NS records are indispensable for a zone cut while accompanying address records and other stray data do not necessarily appear in glue. So we refer to accompanying address records and other stray data as "glue RRs" hereafter. Basically, the NS RR (Resource Record) set indicates the nameserver(s) that be authoritative for a delegated child zone, and the glue records indicate the IP addresse(s) for such name server(s).

The most notable ambiguity arises about glue may be presence of glue RRs in the zone because DNS specifications are vague in this point. In practice, different policies are applied with respect to provisioning glue RRs with the delegating zone [11].

A narrowly defined glue policy can be found in RFC 1034 [12], where glue RRs are provisioned if and only if the name server resides within or below the delegated (child) zone (that is, within the delegated domain). We call this kind of glue RRs "bailiwick". Fig.1 illustrates an example of delegations of a child zone "foo.com" in its parent zone ("com") file. In Fig. 1, line 1 shows a name server "ns.foo.com" residing within the delegated (child) zone "foo.com". Fig. 2 illustrates the zone file of "foo.com", where a child zone "foo.foo.com" is delegated. So line 2 in Fig. 1 shows a name server "ns.foo.foo.com" residing below the delegated (child) zone "foo.com" (that is, within the grandchild zone "foo.foo.com"). So the narrow policy only allows for line 5 and line 6 as the glue RRs without offering line 7 and 8.

| $ORIGIN com. | | | |
|---|---|---|---|
| | Name | Type | Value |
| 1 | foo | NS | ns.foo |
| 2 | foo | NS | ns.foo.foo |
| 3 | foo | NS | ns.foo.net. |
| 4 | foo | NS | ns.bar |
| 5 | ns.foo | A | 192.0.1.1 |
| 6 | ns.foo.foo | A | 192.0.2.2 |
| 7 | ns.foo.net. | A | 192.0.2.2 |
| 8 | ns.bar | A | 192.0.3.3 |
| 9 | bar | NS | ns.foo.net. |

Fig. 1. Example zone data of "com" using RFC 1035 notation [15].

RFC 973 suggests that glue RRs are always provisioned for all name servers [13]. In Fig. 1, this mandatory policy is equivalent to present line 5-8 simultaneously.

A moderate policy is suggested in RFC 1033 and used for the root zone [14]. It suggests that glue RRs are provisioned if and only if the name server resides below the delegating (parent) zone. This policy imposes less constraints on glue RRs than that defined by RFC 1034, allowing for glue RRs of name servers below apex of the parent zone but beyond the bailiwick of the delegated (child) zone. In Fig. 1, the moderate policy presents line 5, 6 and 8 as the glue RRs.

| $ORIGIN foo.com. | | | |
|---|---|---|---|
| | Name | Type | Value |
| 1 | @ | NS | ns |
| 2 | @ | NS | ns.foo |
| 3 | @ | NS | ns.foo.net. |
| 4 | @ | NS | ns.bar.com |
| 5 | ns | A | 192.0.1.1 |
| 6 | foo | NS | ns.foo.org |

Fig. 2. Example zone data of "foo.com" using RFC 1035 notation [15].

*B. The Availability Implications of Glue*

To clarify the definition of glue, we derive the minimum glue RRs required for domain name resolution. And we also discuss the implications and perils of aggressive usage of glue RRs.

*a) The minimum glue RRs required for name resolution*

We can see in Fig. 1 that in-bailiwick glue RR should always be given in the domain. The reason a glue record must exist for nameservers within the domain can be illustrated by looking at what would happen if the glue record were not present. If the query to the "com" nameserver for "foo.com" returned the name but not the IP address of "ns.foo.com", then a further query should be required for the A record of "ns.foo.com"; but since the IP of the "foo.com" nameserver is not yet known, it must query the "com" nameserver, which answers again with the name but not the IP... and so on ad infinitum. So such in-bailiwick glue RRs are among the minimum glue RRs, therefore should be treated as mandatory.

One may consider any out-of-bailiwick glue RR as non-mandatory. However, absence of out-of-bailiwick glue RR may cause resolution unavailability problem because cyclic dependency can be formed based on domain name interdependence. Such cyclic dependency emerges when two or more zones' glue RRs depend on each other in a circular way: to resolve a name server of a glue RR in zone Z1, one needs to first resolve it in Z2, and the name server' zone in Z2 is in turn delegated to a name server below Z3, and so on, until finally the chained delegation interdependency comes back to Z1.

| $ORIGIN com. | | | |
|---|---|---|---|
| | Name | Type | Value |
| 1 | foo | NS | ns.foo.net. |
| 2 | ns.foo.net | A | 192.0.1.1 |
| $ORIGIN net. | | | |
| | Name | Type | Value |
| 1 | foo | NS | ns.foo.com. |
| 2 | ns.foo.com | A | 192.0.2.2 |

Fig. 3. Example zone data of two-zone cyclic dependency using RFC 1035 notation [15].

Fig. 3 shows two examples of cyclic zone dependencies. The first example involves two zone interdependency, where "foo.com" is delegated to an out-of-bailiwick name server "ns.foo.net" in its parent zone "com", and likewise, "foo.net" is in turn delegated to an out-of-bailiwick name server "ns.foo.com" in its parent zone "net". If non-bailiwick glue RR are provided in neither zones, resolution of any name below "foo.com" or "foo.net" will fall into the cyclic zone dependencies failing to find the final answer. A more complicated three zone interdependency is illustrated in Fig. 4, where "com", "net", "org", and "com" are chained in order via cyclic delegated name servers. We can see that glue RR should also be mandatory in case of cyclic zone dependencies to ensure the resolution availability.

| $ORIGIN com. | | | |
|---|---|---|---|
| | Name | Type | Value |
| 1 | foo | NS | ns.foo.net. |
| 2 | ns.foo.net | A | 192.0.1.1 |
| $ORIGIN net. | | | |
| | Name | Type | Value |
| 1 | foo | NS | ns.foo.org. |
| 2 | ns.foo.org | A | 192.0.2.2 |
| $ORIGIN org. | | | |
| | Name | Type | Value |
| 1 | foo | NS | ns.foo.com. |
| 2 | ns.foo.com | A | 192.0.3.3 |

Fig. 4. Example zone data of three-zone cyclic dependency using RFC 1035 notation [15].

To break cyclic zone dependencies without changing any delegation (NS RR), glue RR should be added to at least one zone involved in one cyclic zone dependencies. The zone accompanying with glue RR acts as the door to the authoritative data and may be relied on by any zone involved in the cyclic zone dependencies following the delegation chain.

To summarize, the minimum glue RRs required for domain name resolution consists of in-bailiwick glue RRs and glue RRs in one zone involved in each cyclic zone dependencies.

*b) The implications and perils of aggressive usage of glue RRs*

The aggressive use of glue RRs may cause concerns about the consistency with their authoritative counterparts. For the sake of resolution correctness, the glue RRs in their apex zone should be kept as replicas of the IP address RRs pertaining to the same NS RRs in their authoritative zone. But DNS specifications neither force a tight coupling between the two nor provide any mechanism to ensure such coupling. So in DNS operations, inconsistency problem possibly arises for glue RRs. e.g., a set of IP address RRs are kept both at the authoritative zone A and the apex zone B. Each time the operator of zone A makes changes to its IP address RRs, the operator of zone B must perceive such changes in time (or the operator of zone A actively notifies the operator of zone B about such changes) and update its glue RRs accordingly.

When changes at zone A are not reflected to the glue RRs at zone B, a lame delegation may occur because an IP address of a DNS server that is listed as an authoritative server for a delegated zone cannot provide authoritative answers for names in that zone. Given the potential risks of glue RR inconsistency, excessive glue RRs widely used for delegations may increase the possibility of lame delegations.

### III. THE SECURITY IMPLICATIONS OF GLUE

#### A. Cache Poisoning Attacks

Dan Kaminsky discovered a serious vulnerability in DNS [16], which allows an attacker to spoof DNS records and redirect network clients to alternate servers. The Kaminsky attacks do not require the ability of reading and writing the incoming and outgoing DNS messages of authoritative servers. Instead, a Kaminsky attacker issues a flurry of queries, each for a different random name under the target main domain, aiming at evading the cache of a victim recursive resolver and triggering its outstanding outgoing queries. Such flurry of queries creates many outstanding queries at the victim recursive resolver in the window of responding delays. The attacker then generates replies for each random name query before the real reply arrives from the authoritative server. Each reply guesses the source port and transaction ID of its query to achieve a match, and contains spoofed NS RRs and glue RRs as delegation information. The chance of successful match is very likely in a short time using automatic tools. The recursive resolver will then accept the reply and cache the spoofed NS RRs and glue RRs, which are then utilized as authoritative servers for the afterwards name resolutions.

Man-in-the-Middle attacks are more powerful than Kaminsky attacks in that the MitM attackers have read and write access to network packets belonging to the victim. So the MitM attackers do not have to guess the source port and transaction ID of query to get a match to have a spoofed DNS reply accepted by the recursive resolver. They can simply copy the correct source port and transaction ID in the spoofed DNS reply, which will then poison the cache of the recursive resolver.

Kaminsky and MitM attacks share the ability to inject spoofed DNS data into cache of recursive resolvers. However, the seemingly matching response to the outstanding queries does not necessarily comply with other constrains imposed by DNS specifications and practices. So it is still possible for resolvers to detect some (if not all) deliberately and maliciously formed DNS replies if they are not in compliance with DNS specifications and practices. Resolvers may perform additional checking on the received matching response before those pass the checking are finally accepted.

#### B. Bailiwick-checking and Its Improvements

The most prevalent practice of thwarting DNS tampering attacks is to implement bailiwick-checking. Basically, the purpose of the bailiwick-checking is to prevent malicious response from providing out-of-bailiwick DNS records as part of a referral. In history, the bailiwick-checking was motivated by kashpureff cache poisoning [5]. In 1997, Eugene Kashpureff of Alternic poisoned multiple DNS caching servers which later redirected web traffic from internic.net to alternic.com. The hack showed incorrect NS or A records provided in glue could be trusted by some DNS implementations and cached. This vulnerability produces bailiwick-checking for both NS and glue records to fix it.

While DNS specifications does not define a concrete bailiwick rule, some implementations, such as BIND [17], Unbound [18], and MaraDNS [19], adopt their characteristic bailiwick-checking algorithms. Here we analyze bailiwick-checking algorithms of the three implementations and explore what they mean for the MitM attacks.

##### a) BIND

BIND's bailiwick-checking consists of NS checking and glue checking.

For the NS records, the resolver checks whether the domain in question is a subdomain of the NS records in the authority section of the received response. If it is, the resolver next checks whether the domain in the authority section is a subdomain of the current domain where the recursive DNS resolution resides. Only if both conditions hold, the resolver caches the NS-type RRset received in the referral. The NS checking ensures only in-bailiwick NS records in the referral are accepted and used by resolvers for replying to their clients.

For the glue records, the next step is to determine whether to cache them in the additional section of the referral. If the domain of glue record in the additional section is a subdomain of currently residing domain, the glue record is cached; otherwise, it is not cached and the resolver will initiate new queries for the domain of glue record in the additional section.

##### b) Unbound

Unbound's bailiwick checking is very different from BIND. All records, that are pertinent to the domains out of the bailiwick of the current domain where the recursive DNS resolution resides, are simply removed from the received responses. The remaining records in the additional and authority sections, if pass the NS checking and glue checking, are cached, but, by default, not sent to clients. The most remarkable difference of the bailiwick checking between Unbound and BIND is that Unbound simply denies out-of-bailiwick domains in the authority section as referral while BIND accepts them by issuing new queries for them. Thus BIND is more flexible and resilience in that at least one in-bailiwick referral is a must for Unbound and BIND has no such constraints.

##### c) MaraDNS

The bailiwick logic is weakened by MaraDNS. It does not cache the authority and additional section of responses, and refrains from performing much of bailiwick-checking. Furthermore, even for referral responses, MaraDNS caches neither the NS mapping from the domain name to an authoritative server name (authority section), nor the A mapping from the latter name to an IP address (additional section). Instead, MaraDNS simply stores an authority section with a mapping from the domain name to the IP address. The bailiwick-checking here is very loose: no matter whether the domains in referral is in-bailiwick or out-of-bailiwick, and no matter where the additional section contains glue records or authoritative records, just accept all domain-to-IP mappings that are provided. So such mappings may deviate from the authoritative and more trustworthy ones.

*d) Improvement Proposals*

Note that NS checking may add no security profit to DNSSEC-aware resolvers. Because signature over DS records itself secure the authentication of the NS records. In other words, no attacker can forge a NS record in the referral, and meanwhile, get the forged verified.

There is subtle misconception and misbehavior for BIND's glue record checking. A subdomain of currently residing domain by no means guarantees it falls into the bailiwick of the currently residing domain. Instead, a subdomain of currently residing domain may be delegated into a new zone, which is obviously out of the bailiwick of the currently residing domain. e.g., ns.example.com is a subdomain of com, and also a subdomain of example.com when example.com is delegated by com. In that case, the glue records provided from com for ns.example.com are not authoritative ones, whereas the authoritative ones are indicated nowhere but in example.com. So BIND's bailiwick-checking risks taking the glue records as trustworthy and caching them as well as losing the change to fetch the authoritative ones. Under BIND's bailiwick-checking algorithm, the glue records should be carefully configured in com as a replica of the authoritative ones maintained elsewhere. Any inconsistency between them may impact domain availability or cause lame delegation.

To learn from lessons of DNS implementations in use and clarify the best current practice, we propose that at least two principles for bailiwick-checking should be adhered:

● An authoritative server name that a domain is delegated to, either in-bailiwick or out-of-bailiwick, should always be allowed by bailiwick-checking. This fosters a diversity of DNS delegations.

● Any glue records pertinent to the authoritative server name of delegation may be temporarily used to fetch the authoritative counterparts. Then the authoritative counterparts replace the corresponding glue records and get cached by resolvers. This ensures only authoritative mappings are cached. As we know, glue RRs can never be part of the zone's authoritative data. They are ranked as the least trustworthiness in data from all sources [20]. Moreover, they would never be returned as answers to a received query by a resolver. To ensure trustworthiness of glue RRs, a trustworthiness validation using data from glue RRs' real authorities is required even if glue RRs are temporally used during a resolver's domain name resolution. e.g., each of the glue RRs in line 5-8 of Fig. 1 should be validated by resolvers when they are fetched from the parent zone "com" as referrals and then directly used to reach the child zone "foo.com". The so called validation means that the resolver should solicit the authoritative counterparts of the glue RRs from their authoritative servers, and then check whether the previously obtained glue RRs are consistent with the later obtained authoritative counterparts. If inconsistency occurs, the latter should be regarded as more trustworthy and replace the former. As examples, the consistency checking of validation takes place between line 5 in Fig. 1 (glue RR) and line 5 in Fig. 2 (authoritative RR), line 6 in Fig. 1 (glue RR) and the authoritative RR in the name server "ns.foo.org" (see line 6 in Fig. 2), line 7 in Fig. 1 (glue RR) and the authoritative RR in the name server of "ns.foo.net", and so on.

## C. Failure of Bailiwick-checking against Cache Poisoning

Bailiwick-checking is not always effective for some deliberately and meticulously orchestrated attacks especially in DNSSEC oblivious occasions. In this section, we discuss how cache poisoning attackers can successfully pass the proposed version of bailiwick-checking in Section B and redirect resolvers to a bogus nameserver in their control. Here we assume that DNSSEC is not deployed to protect DNS data.

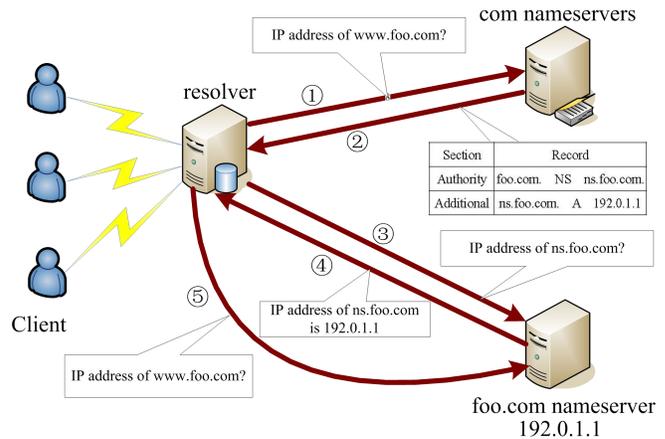

Fig. 5. Example name resolution using the improved bailiwick-checking.

An example name resolution using the improved bailiwick-checking is illustrated in Fig. 5. The detailed process is elaborated in the following steps:

① The resolver queries the IP address of "www.foo.com" to the "com" nameservers;

② The "com" nameservers return a referral to resolver. The referral indicates that "foo.com" is delegate to a nameserver "ns.foo.com" (in the authority section), and the IP address of "ns.foo.com" is "192.0.1.1" (as a glue RR in the additional section);

③ The resolver performs bailiwick-checking for the glue RR: first it temporarily accepts the glue RR and follows it to query the authoritative version of the glue RR to "192.0.1.1" ;

④ When the nameserver "192.0.1.1" returns an authoritative response confirming that the IP address of "ns.foo.com" is "192.0.1.1", the resolver successfully completes its bailiwick-checking and trusts the delegation and its glue RR;

⑤ The resolver queries the IP address of "www.foo.com" to the "foo.com" nameserver.

The improved bailiwick-checking can only guarantee the consistency between the glue RR and its claimed authoritative counterpart. But cache poisoning attacker can maintain the same authoritative data as the bogus glue RR in its bogus nameservers, and redirect bailiwick-checking to it by injecting the bogus glue RR. To compare with the similar scenario in Fig. 5, such attacks are illustrated in Fig. 6. The detailed process is elaborated in the following steps:

① The resolver queries the IP address of "www.foo.com" to the "com" nameservers, and the query is either initiated by Kaminsky attackers or overheard by MitM attackers;

② The cache poisoning attackers manage to inject a referral to resolver. The referral may keep the nameserver as "ns.foo.com" (in the authority section), but spoof the IP address of "ns.foo.com" as "192.0.2.2" (as a glue RR in the additional section);

③ The resolver performs bailiwick-checking by querying the IP address of "ns.foo.com" to the bogus nameserver "192.0.2.2" ;

④ The bogus "foo.com" nameserver configures an A type RR in its zone indicating that the IP address of "ns.foo.com" is "192.0.2.2". So it response with the RR to the resolver, so the resolver successfully completes its bailiwick-checking and trusts the delegation and its bogus glue RR;

⑤ The resolver queries the IP address of "www.foo.com" to the bogus "foo.com" nameserver, so the response may be also bogus at the mercy of the attackers.

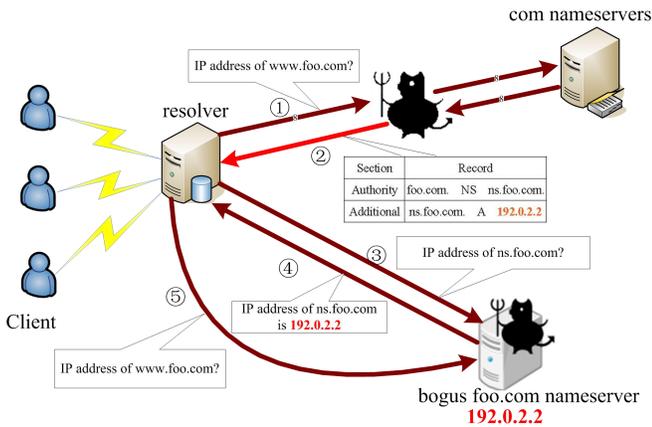

Fig. 6. Example cache poisoning against bailiwick-checking.

### D. DNSSEC and Its Defending against Cache Poisoning

#### a) DNSSEC

Domain Name System Security Extensions (DNSSEC) standards [8], [9] was designed to provide data integrity, origin authenticity, and authenticated denial of existence based on public key cryptography.

Public/private key pairs are used for the authentication of each zone. The public keys are stored in DNSKEY RRset, and all the signatures signed over other records using DNSSKEY are stored in RRSIG RRset. Both DNSKEY and RRSIG are included in the zone as part of it. In response to a query soliciting DNSSEC, an authoritative server returns both the requested data and its associated RRSIG RRset. A resolver that has learned the DNSKEY of the requested zone can verify the origin authenticity and integrity of the reply data using the DNSKEY. To resist replay attacks, each signature carries a definitive expiration time.

In order to authenticate the DNSKEY for a given zone, say "www.foo.com", the resolver needs to construct a chain of trust that follows the DNS hierarchy from a trusted root zone key down to the key of the zone in question. In the ideal case, the public key of the DNS root zone would be obtained offline in a secure way and stored at the resolver, so that the resolver can use it to authenticate the public key of TLD "com"; the public key of "com" would then be used to authenticate the public key of its child zone "foo.com". A parent zone must encode the authentication of each of its child zone's public keys in the DNS. To accomplish this, the parent zone creates and signs a Delegation Signer (DS) RR that corresponds to a DNSKEY RR at the child zone, and creates an authentication link from the parent to child. It is the child zone's responsibility to request an update to the DS RR every time the child's DNSKEY changes.

When a DNS request queries a non-existing record, an authoritative server returns a negative NXDOMAIN or NODATA response, where NXDOMAIN means the queried name is non-existent and NODATA means the queried RR type of an existent name is non-existent. To authenticate such responses, DNSSEC introduces new types of records, namely NSEC and NSEC3, to be signed and thus authenticated.

NSEC [8] allows to cryptographically prove that a RRset does not exist by spanning a gap between two domain names in a zone. NSEC specifies what type of records exist at a name where it resides and points to the next domain name (in canonical order) in the zone. One obvious drawback of NSEC is that it makes it possible for enumerating the entire zone by walking along the NSEC chain from one name to the next. To fix the zone data exposure risk, an alternative was proposed: NSEC3 [21] specifies hashed names, not original names, at both ends of a gap. In the response to a query soliciting DNSSEC authenticated non-existing record, an authoritative server responses with NSEC/NSEC3 records along with their signatures.

#### b) DNSSEC's Defending against Cache Poisoning

We analyze how DNSSEC is in place to protect DNS RRs from cache poisoning attacks:

● If a zone is signed, then all its authoritative RRsets are signed by DNSSEC and resistant against cache poisoning attacks. This is due to the incapability of the attackers to produce any signatures signed using the authenticated DNSKEY over the forged authoritative RRsets. Note that the NS RRsets at the delegating points and their glue RRs are not among the zone's authoritative RRsets. And they are not signed by DNSSEC.

● If a zone is signed and its child zone is signed, then its NS RRSet at the delegating point is resistant against cache poisoning attacks. As specified by the DNSSEC protocol, the signed child zone has its DS RR in the parent zone, and the DS is signed and thus verifiable using the authenticated DNSKEY of the parent zone. The DS RR, which is the digest of the child zone's DNSKEY, provides secure link to the child zone's DNSKEY. In the child zone, its DNSKEY is used to sign its authoritative RRSets, of course, including its NS RRSet. In some practices, DNSKEY RRset is split into KSK (Key Signing Key) and ZSK (Zone Signing Key). KSK is used to sign the DNSKEY RRset and ZSK is used to sign the entire zone except the DNSKEY RRset. The DS RR is mapped into the KSK, which is used to sign the ZSK. And ZSK is used to sign its NS RRset. The trust link built between the parent zone and the child zone is shown in Fig. 7.

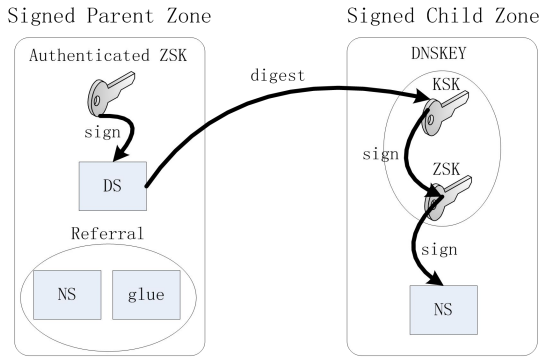

Fig. 7. The trust link built between the parent zone and the child zone.

While the NS RRsets at the delegating points are not signed by DNSSEC, any attempt to tamper them in the referral response should also provide the bogus NS RRset in the child (authoritative) zone as the replica of the bogus referral. The former is signed by the ZSK of the child zone, which is in turn verified through the parent-child trust link (see Fig. 7). So the attacker is unable to form the signature over the bogus NS RRset using the genuine ZSK.

● The glue records in a signed zone are vulnerable to cache poisoning attacks if and only if their owner names falls into an unsigned zone. The glue records are expected to be consistent with their authoritative versions in the bailiwick-checking. The attackers can always tamper the glue records and meanwhile tamper their authoritative versions into the same bogus ones if both of them are not signed. However, when the owner names of the glue records are authenticated by DNSSEC and thus tamper-resistant, any attempts to tamper the glue records will cause inconsistency between the glue records and their authoritative versions. Such inconsistency will be detected and prevented from being accepted by the bailiwick-checking. An authoritative server name that a domain is delegated to, either in-bailiwick or out-of-bailiwick, should always be allowed by bailiwick-checking. This fosters a diversity of DNS delegations.

*c) DNSSEC-aware Bailiwick-checking*

As discussed above, DNSSEC strengthens the capability of defending against cache poisoning attacks. And such strengthening exploits DNSSEC-aware bailiwick-checking, which integrates DNSSEC verification into the original DNSSEC-oblivious bailiwick-checking:

● If a zone is signed, then all glue records in it should pass the glue verification before they can be used to reach the child zone. For any glue records whose owner names are in the signed zone, the glue verification is initiated as the standard DNSSEC verification over the glue records. That is, fetch the authoritative versions of the glue records and then verify them per DNSSEC specifications.

● If a zone is signed and its child zone is signed, the bailiwick-checking should verify the authoritative NS RRset of the child zone before comparing it with the NS RRsets at the delegating point. Similar to the original DNSSEC-oblivious bailiwick-checking, the DNSSEC-aware bailiwick-checking should also follow the NS RRsets at the delegating point as well as the glue records to obtain the authoritative NS RRset from the child zone. Note that the glue record, if any, should be verified first through the glue checking procedure stated above. The DNSSEC-aware bailiwick-checking also obtains and verifies the ZSK of the child zone following the trust of chain (see Fig. 7). Using the authenticated ZSK, the bailiwick-checking checks the authenticity of the authoritative NS RRset. Failure of such authentication indicates the delegation under check is untrustworthy. Otherwise, the verified authoritative NS RRset is then compared against the NS RRsets at the delegating point, and consistency between them is accepted as trustworthy.

*E. DNSSEC Vulnerabilities and the Exploit*

As glue records are not signed, the first concern about DNSSEC vulnerabilities lies in the forgeability of glue records. The second class of venerabilities comes from the limited DNSSEC deployment and the NSEC's "Opt-Out" option allowing for insecure delegations to unsigned zones.

*a) Glue Tampering Attacks*

We discuss whether glue RR is tamper-resistant with DNSSEC. While glue records are not signed in the signed zone where it resides as glue records, the proposed DNSSEC-aware bailiwick-checking can validate them only if they are signed at their authoritative zones. However, we will illustrate in the next section that in some cases, perhaps corner ones, a validated signed response does not necessarily indicate glue record's trustworthiness. We call this kind of zones as "signed insecure zone".

The relative location relationship in the DNS hierarchy between glue records' residing zone and authoritative zone may be diverse. We make classified discussions as the following four classes:

*1) The authoritative zone falls beyond the domain of the residing zone.*

In the proposed DNSSEC-aware bailiwick-checking, the recursive resolver relies on the standard name resolution of the glue RR to find the authoritative version and its signature. But the recursive resolver cannot foretell whether the authoritative zone of the glue RR is signed or not until the glue RR is resolved, because there is no RR as indicator in its residing zone. So it is possible for the attackers to forge DNSSEC-oblivious authoritative response without arousing any recursive resolver's explicit suspicion about the authenticity of the bogus response. We will show in the next section that if the authoritative zone of the glue RR satisfies "signed insecure zone", the attackers may redirect the glue RR to an unsigned child zone of it, and then the finally answer is vulnerable to tamping attacks.

We also consider the possibility of tampering the DNSSEC signed authoritative response for the glue RR. This is almost equivalent to compromising the private key or the entire authoritative zone of the glue RR. DNSSEC protocols themselves provide sufficient protection against such compromising.

To summarize, the following three cases are listed:

● The authoritative zone is unsigned. Glue records are vulnerable to tamping attacks.

● The authoritative zone is signed but "signed insecure zone". Glue records are vulnerable to tamping attacks.

● The authoritative zone is signed and not "signed insecure zone". Glue records are tamper-resistant.

*2) The authoritative zone lies below the domain of the residing zone but not within any of the residing zone's child zone.*

The glue records' residing zone itself is the authority for them and thus signs them. So the glue records are protected by DNSSEC against tamping attacks.

*3) The authoritative zone lies below the domain of the residing zone and within one of the residing zone's child zone other than its child zone pertaining to the glue record's owner name.*

The vulnerability analysis is identical to Case 1. It is obvious that the unsigned child zone implicates the unverifiability of the authoritative version because the recursive resolver can never fetch and check its signature. However, it is notable that the signed child zone does not necessarily mean verifiability of the authoritative version. Given that the authoritative version lies within the child zone, the child zone does not have to be authoritative for it. It is commonplace that the authoritative version is delegated to a child zone of the child zone, or its grandchild zone, etc. While the child zone is signed, there is still possibility that its child zone with the authoritative version below it is unsigned and unverifiable.

*4) The authoritative zone lies below the domain of the residing zone and within the residing zone's child pertaining to the glue record's owner name.*

As the residing zone's child pertaining to the glue record's owner name is signed, the vulnerability analysis is simplified from Case 1 as:

● The authoritative zone is signed but "signed insecure zone". Glue records are vulnerable to tamping attacks.

● The authoritative zone is signed and not "signed insecure zone". Glue records are tamper-resistant.

*b) Glue Redirection Attacks*

Attackers may successfully tamper the glue records by passing the bailiwick-checking in certain cases as analyzed above. Then all queries towards the victim domain will be redirected to the malicious IP addresses indicated by the tampered glue records. The impacts are three-fold:

● The real authoritative nameservers of the victim domain are prevented from receiving any query traffic towards the victim zone.

● All query traffic towards the victim domain are gathered and abused by the attackers, who may take full control of the malicious IP addresses indicated by the tampered glue records. Such concerns may be about the privacy and sensitive information about Internet users leaking from the aggregated DNS query data analysis.

● The faked authoritative nameservers for the signed victim zone may response in any way to the queries: ① using the real authoritative data, which are obtained from the real authoritative server either by zone transfer (if permitted) or by on demand queries; ② using the fabricated data, which are likely to result in failure of validation at the DNSSEC-aware recursive resolvers or validators (so the answer is labeled as bogus) ; ③ using the fabricated but vulnerability-exploiting data, which are likely to be validated by the DNSSEC-aware recursive resolvers or validators and be trusted by them (so the answer is labeled as trustworthy); ④ just playing as unresponsive at all, which is kind of DoS (Denial of Service) attacks.

Among the four classes of possible behaviors taken by the faked authoritative nameserver, the 3rd is the most hazardous one because it can potentially cheat the DNSSEC validators into accepting the bogus data as authenticated. Such attacks are called "glue redirection attacks" hereafter.

*c) NSEC3 Opt-Out Vulnerabilities and the Exploit*

Opt-Out is normally relevant as a feature of NSEC3 when a zone has a significant number of delegations (frequently called a delegation-centric zone in the jargon) and inhibits generation of NSEC3 RRs when an unsigned delegation occurs. An unsigned delegation has NS and glue records but no DS RR(s). In an Opt-Out zone, hashed owner names of unsigned delegations may be excluded from the NSEC3 chain. An NSEC3 resource record that has the Opt-Out flag set to 1 may have its span cover the hash of an owner name or "next closer" name of an unsigned delegation. Such a relatively modest step can have a huge impact on zone file size as well as cost of maintaining NSEC3 chain for frequently updated insecure delegations especially, but not exclusively, for TLD zones.

In essence, NSEC3 Opt-Out is a transition method to ease the operation of delegation-centric zone when DNSSEC is not so prevalent. Although Opt-Out is advantageous in terms of its manageability for a diminished secure zone, it also blurs the authenticated presence of different responses by providing somewhat overlapping NSEC3 records for them. This introduces the vulnerabilities of secure shift from the real response to an forged alternative, and such shift is unlikely to be detected and prevented by the current DNSSEC specifications. We present here three classes of Opt-Out vulnerabilities:

*1) Transforming an authenticated positive response to a insecure delegation.*

An authenticated positive response is illustrated as an example in Fig. 8, where the domain in question is "a.c.example". The authenticated positive response can be transformed by attackers into a insecure delegation (see Fig. 9) if and only if the following conditions are satisfied in order:

① At least one subdomain (e.g., "c.example" in Fig. 9) of the victim domain (e.g., "a.c.example" in Fig. 9) is non-existent. Otherwise, the transformation is impossible.

② The hash value of the non-existent subdomain (e.g., "c.example" in Fig. 9) is covered by a NSEC3 Opt-Out RR (e.g., "35mthgpgcu..." NSEC3 RR in Fig. 9). And neither the NSEC3 RR's owner name nor its next name match the hash value of the victim domain. That is (in Fig. 8)

H(a.c.example) ≠ 35mthgpgcu1qg68fab165klnsnk3dpvl

and

H(a.c.example) ≠ b4um86eghhds6nea196smvmlo4ors995

Otherwise, the transformation is impossible.

③ There is a NSEC3 RR (e.g., "0p9mha..." NSEC3 RR in Fig. 8) that matches the closest encloser (e.g., "example" in Fig. 8). And the NSEC3 RR's next name does not match the hash value of the victim domain. That is (in Fig. 8)

H(a.c.example) ≠ 2t7b4g4vsa5smi47k61mv5bv1a22bojr

Otherwise, the transformation is impossible.

The RR on which such authenticated transformation takes place is called "signed insecure RR". A zone with at least one "signed insecure RR" is the aforementioned "signed insecure zone".

The first condition decides whether the victim domain falls into a child zone, which may be insecurely delegated potentially covered by an NSEC3 Opt-Out RR. The second condition identifies the existence of the NSEC3 Opt-Out RR covering the subdomain, and its two mismatch checking ensure that the victim domain is not accidentally leaked from the owner name or the next name of the NSEC3 Opt-Out RR. The third condition ensure that the victim domain is not accidentally leaked from the next name of the closest encloser of NSEC3 RR.

We use the following notations for a general case: the target zone: DOMAIN; the victim domain: $A_n$. $A_{n-1}$. ... . $A_2$. $A_1$. DOMAIN. The attack model is elaborated in the following steps:

① Once finding that the response from "DOMAIN" for "$A_n$. $A_{n-1}$. ... . $A_2$. $A_1$. DOMAIN" is a positive response, the attacker initiates the suffix to check as the victim domain, and then moves to step 2.

② The attacker moves the current suffix one label to the left as the updated suffix to check. If the suffix to check is "DOMAIN", return with an indicator of "inconvertibility" for the victim domain. Otherwise, moves to step 3.

③ The attacker sends the query for the suffix to the real authoritative nameservers of "DOMAIN" . If the response is positive, go to step 2. Otherwise, extract the covering NSEC3 RR and the closest encloser NSEC3 RR in the authenticated negative response, and then check whether Condition 2 and 3 are both satisfied. If so, return with an indicator of "convertibility" for the victim domain. Otherwise, go to step 2.

④ If the victim domain is labeled as "convertibility", forge the transformed response as follows:

Add the delegation whose owner name is the victim domain's current suffix, and whose nameserver(s) and glue RR(s) are arbitrary, perhaps malicious.

Add the covering NSEC3 RR and the closest encloser NSEC3 RR as well as their signatures extracted from the real authenticated negative response.

Add other RRs if necessary.

```
;; Question
  a.c.example.    IN   A

;; Answer
  a.c.example.  IN  A   192.0.2.1
  a.c.example.  RRSIG  A 5 3 3600 20040509183619 (
                 20040409183619 38519 example.
                 Il2WTZ+Bkv+OytBx4LItNW5mjB4RCwhOO8y1
                 XzPHZmZUTVYL7LaA63f6T9ysVBzJRI3KRjAP
                 H3U1qaYnDoN1DrWqmi9RJe4FoObkbcdm7P3I

;; Authority
  example.   NS    ns1.example.
  example.   NS    ns2.example.
  example.   RRSIG  NS 5 1 3600 20040509183619 (
                 20040409183619 38519 example.
                 gl13F00f2U0R+SWiXXLHwsMY+qStYy5k6zfd
                 EuivWc+wd1fmbNCyql0Tk7lHTX6UOxc8AgNf
                 4ISFve8XqF4q+o9qlnqIzmppU3LiNeKT4FZ8
                 RO5urFOvoMRTbQxW3U0hXWuggE4g3ZpsHv48
                 0HjMeRaZB/FRPGfJPajngcq6Kwg= )

;; Additional
  ...
```

Fig. 8. Example of an authenticated positive response.

*2) Transforming a non-existent negative response to an insecure delegation.*

An authenticated non-existent negative response is illustrated as an example in Fig. 10, where the domain in question is "a.c.x.w.example". The authenticated non-existent negative response can be transformed by attackers into a insecure delegation (see Fig. 11) if and only if the following condition is satisfied: the non-existent domain is covered by a NSEC3 Opt-Out RR.

```
;; Question
a.c.example.     IN    A

;; Answer
;; (empty)

;; Authority
c.example.   NS   ns1.c.example.
             NS   ns2.c.example.

;; NSEC3 Opt-Out RR that covers the "next closer" name (c.example)
;; H(c.example) = 4g6p9u5gvfshp30pqecj98b3maqbn1ck

35mthgpgcu1qg68fab165klnsnk3dpvl.example. NSEC3 1 1 12 aabbccdd (
            b4um86eghhds6nea196smvmlo4ors995 NS DS RRSIG )
35mthgpgcu1qg68fab165klnsnk3dpvl.example. RRSIG NSEC3 7 2 3600 (
            20150420235959 20051021000000 40430 example.
            g6jPUUpduAJKRljUsN8gB4UagAX0NxY9shwQ
            Aynzo8EUWH+z6hEIBlUTPGj15eZll6VhQqgZ
            XtAIR3chwgW+SA== )

;; NSEC3 RR that matches the closest encloser (example)
;; H(example) = 0p9mhaveqvm6t7vbl5lop2u3t2rp3tom

0p9mhaveqvm6t7vbl5lop2u3t2rp3tom.example. NSEC3 1 1 12 aabbccdd (
            2t7b4g4vsa5smi47k61mv5bv1a22bojr MX DNSKEY NS
            SOA NSEC3PARAM RRSIG )
0p9mhaveqvm6t7vbl5lop2u3t2rp3tom.example. RRSIG NSEC3 7 2 3600 (
            20150420235959 20051021000000 40430 example.
            OSgWSm26B+cS+dDL8b5QrWr/dEWhtCsKlwKL
            IBHYH6blRxK9rC0bMJPwQ4mLIuw85H2EY762
            BOCXJZMnpuwhpA== )

;; Additional
ns1.c.example.  A    192.0.2.7
ns2.c.example.  A    192.0.2.8
```

Fig. 9. Example of a NSEC3 Opt-Out insecure delegation response.

The attack model is to manipulate the response as follows:

Take out the SOA RR and the NSEC3 RR that covers wildcard at the closest encloser as well as their associated RRSIG RRs.

Add the delegation whose owner name is the "next closer" name of the victim non-existent domain, and whose nameserver(s) and glue RR(s) are arbitrary, perhaps malicious.

```
;; Question
a.c.x.w.example.     IN A

;; Answer
;; (empty)

;; Authority
example.    SOA   ns1.example. bugs.x.w.example. 1 3600 300 (
            3600000 3600 )
example.    RRSIG  SOA 7 1 3600 20150420235959 20051021000000 (
            40430 example.
            Hu25UIyNPmvPIVBrldN+9Mlp9Zql39qaUd8i
            q4ZLlYWfUUbbAS41pG+68z81q1xhkYAcEyHd
            VI2LmKusbZsT0Q== )

;; NSEC3 Opt-Out RR that covers the "next closer" name (c.x.w.example)
;; H(c.x.w.example) = 0va5bpr2ou0vk0lbqeeljri88laipsfh

0p9mhaveqvm6t7vbl5lop2u3t2rp3tom.example. NSEC3 1 1 12 aabbccdd (
            2t7b4g4vsa5smi47k61mv5bv1a22bojr MX DNSKEY NS
            SOA NSEC3PARAM RRSIG )
0p9mhaveqvm6t7vbl5lop2u3t2rp3tom.example. RRSIG NSEC3 7 2 3600 (
            20150420235959 20051021000000 40430 example.
            OSgWSm26B+cS+dDL8b5QrWr/dEWhtCsKlwKL
            IBHYH6blRxK9rC0bMJPwQ4mLIuw85H2EY762
            BOCXJZMnpuwhpA== )

;; NSEC3 RR that matches the closest encloser (x.w.example)
;; H(x.w.example) = b4um86eghhds6nea196smvmlo4ors995

b4um86eghhds6nea196smvmlo4ors995.example. NSEC3 1 1 12 aabbccdd (
            gjeqe526plbf1g8mklp59enfd789njgi MX RRSIG )
b4um86eghhds6nea196smvmlo4ors995.example. RRSIG NSEC3 7 2 3600 (
            20150420235959 20051021000000 40430 example.
            ZkPG3M32lmoHM6pa3D6gZFGB/rhL//Bs3Omh
            5u4m/CUiwtblEVOaAKKZd7S959OeiX43aLX3
            pOv0TSTyiTxIZg== )

;; NSEC3 RR that covers wildcard at the closest encloser (*.x.w.example)
;; H(*.x.w.example) = 92pqneegtaue7pjatc3l3qnk738c6v5m

35mthgpgcu1qg68fab165klnsnk3dpvl.example. NSEC3 1 1 12 aabbccdd (
```

```
                b4um86eghhds6nea196smvmlo4ors995 NS DS RRSIG )
35mthgpgcu1qg68fab165klnsnk3dpvl.example. RRSIG NSEC3 7 2 3600 (
                20150420235959 20051021000000 40430 example.
                g6jPUUpduAJKRljUsN8gB4UagAX0NxY9shwQ
                Aynzo8EUWH+z6hEIBlUTPGj15eZll6VhQqgZ
                XtAIR3chwgW+SA== )

;; Additional
;; (empty)
```

Fig. 10. Example of an authenticated non-existent negative response.

### 3) Transforming an insecure delegation to a non-existent negative response

Conversely, the attackers may transform an insecure delegation to a non-existent negative response. The attack model is to manipulate the response as follows:

To obtain the NSEC3 RR that covers wildcard at the closest encloser (e.g., "*.x.w.example"), the attacker can send a query for a non-existent domain (e.g., "ayh2s0e9s.x.w.example" ) to the real authoritative nameservers of the target zone (e.g., "example"), and then extract the NSEC3 RR in demand from the response as well as its signature.

To obtain the SOA RR, the attacker can send a query for the SOA RR of the target zone (e.g., "example") to its real authoritative nameservers, and then extract the SOA RR in demand from the response as well as its signature.

Taken out the insecure delegation in the answer section and the glue RR(s) in the additional section.

Add the NSEC3 RR that covers wildcard at the closest encloser and the SOA RR as well as their RRSIGs in the authority section.

```
;; Question
a.c.x.w.example.        IN A

;; Answer
;; (empty)

;; Authority
c.x.w.example.   NS    ns1.c.x.w.example.
                 NS    ns2.c.x.w.example.

;; NSEC3 Opt-Out RR that covers the "next closer" name (c.x.w.example)
;; H(c.x.w.example) = 0va5bpr2ou0vk0lbqeeljri88laipsfh

0p9mhaveqvm6t7vbl5lop2u3t2rp3tom.example. NSEC3 1 1 12 aabbccdd (
                2t7b4g4vsa5smi47k61mv5bv1a22bojr MX DNSKEY NS
                SOA NSEC3PARAM RRSIG )
0p9mhaveqvm6t7vbl5lop2u3t2rp3tom.example. RRSIG NSEC3 7 2 3600 (
                20150420235959 20051021000000 40430 example.
                OSgWSm26B+cS+dDL8b5QrWr/dEWhtCsKlwKL
                IBHYH6blRxK9rC0bMJPwQ4mLIuw85H2EY762
                BOCXJZMnpuwhpA== )

;; NSEC3 RR that matches the closest encloser (x.w.example)
;; H(x.w.example) = b4um86eghhds6nea196smvmlo4ors995

b4um86eghhds6nea196smvmlo4ors995.example. NSEC3 1 1 12 aabbccdd (
                gjeqe526plbf1g8mklp59enfd789njgi MX RRSIG )
b4um86eghhds6nea196smvmlo4ors995.example. RRSIG NSEC3 7 2 3600 (
                20150420235959 20051021000000 40430 example.
                ZkPG3M32lmoHM6pa3D6gZFGB/rhL//Bs3Omh
                5u4m/CUiwtblEVOaAKKZd7S959OeiX43aLX3
                pOv0TSTyiTxIZg== )
;; Additional
  ns1.c.x.w.example.    A    192.0.2.7
  ns2.c.x.w.example.    A    192.0.2.8
```

Fig. 11. Example of a NSEC3 Opt-Out insecure delegation response.

## IV. MEASUREMENTS OF GLUE'S AVAILABILITY AND SECURITY

NIST provides the estimation of IPv6 & DNSSEC deployment status by sampling some second level domains [22]. Our measurements borrow the same second level domains list as used in the NIST's estimation. The list has 1070 large industrial US domains, which come from Fortune 1000 list as well as domains collected from the Alexa list of the top 100 sites in the US.

First, we measure the distribution of different glue types as well as glue's redundancy and deficiency. As analyzed in Section II, glue types can be classified as:

1) Self-contained; glue RR(s) must be present to ensure availability.

2) Cyclic dependency; the minimum glue RR(s) to ensure availability are the one(s) only in one zone involved in the cyclic dependency.

3) Out-of-bailiwick; the minimum glue RR(s) to ensure availability are none.

Excessive glue RR(s) except the minimum can be taken as redundancy and give rise to inconsistency concerns. The results of glue type and status distribution are shown in Table 1. Some domains' glue may cover more than one glue types, e.g., a domain may have a self-contained glue and an out-of-bailiwick glue simultaneously. We can see in Table 1 that self-contained glue is the most prevalent glue type of the three, covering more than half of the total domains. Only a small minority of domains use cyclic dependency glue, perhaps due to the complexity of management. In general, the provision of glue RRs are sufficient and far more than necessary in terms of domain resolution availability. Specifically, for self-contained glues, where glue RR(s) should always be present, no volitation is found. And all cyclic dependency glues provide excessive glue RR(s). The most remarkable redundancies lie in out-of-bailiwick glues, of which only 12 domains add no glue RR(s) and the rest are attached with superfluous glue RR(s). In percentage terms, 49.2% of the total are subject to glue RR redundancies, and 98.3% out of the possibly excessive glues have redundant glue RRs.

TABLE I. GLUE TYPE AND STATUS DISTRIBUTION

| Glue Type/Status | Present/Redundant | Minimum | Absence |
|---|---|---|---|
| Self-contained | 728 |  | 0 |
| Cyclic dependency | 12 | 0 | 0 |
| Out-of-bailiwick | 693 |  | 12 |

We then measure the security vulnerabilities of the domains signed by DNSSEC. Of the list of domains, only 17 domains are signed, which are mostly owned by IT companies or organizations. This shares the same conclusion with previous measurements that DNSSEC deployment is quite limited in the domain levels lower than TLD. Except for the IT industry, the Internet society lacks the awareness and readiness of DNSSEC. The low DNSSEC penetration rate may also be attributed to the insufficient technical and/or funding capacities for a large number of non-IT or small companies or organizations.

According to the bailiwick location of glue and the status of DNSSEC deployment, we classify the glues into:

1) Self-contained. The glue falls into the SLD in investigation, and the SLD is, of course, signed.

2) Signed cross-domain. The glue falls beyond the parent(or TLD) of the SLD in investigation and is signed.

3) Unsigned cross-domain. The glue falls beyond the parent(or TLD) of the SLD in investigation and is unsigned.

4) Signed cross-subdomain. The glue falls into a SLD which is other than the SLD in investigation and shares the TLD with it, and is signed.

5) Unsigned cross-subdomain. The glue falls into a SLD which is other than the SLD in investigation and shares the TLD with it, and is unsigned.

As analyzed in Section III, we evaluate the glue and zone vulnerabilities of the domains with DNSSEC. Any domain that has at least one of its glue RR(s) unsigned or one of its signed glue RR(s) have the NSEC3 Opt-Out vulnerabilities, are regarded as glue forgeability. Glue forgeability means the domain is vulnerable to glue tampering attacks. The domains with glue forgeability may be further subject to zone forgeability. Zone forgeability means the domain may be redirected to the bogus nameserver(s), which then hijack its RR(s) in the zone. Zone forgeability also exploits the NSEC3 Opt-Out vulnerabilities. To simplify the evaluations, we do not enumerate the RRs in the zone and test their forgeability one by one. Instead, any domain with glue forgeability which sets the NSEC3 Opt-Out is regarded as zone forgeability.

The bailiwick location of glue and the status of DNSSEC deployment as well as the glue and zone vulnerabilities are shown in Fig. 12. We can see that if the glue falls beyond the parent(or TLD) of the SLD in investigation, whether it is signed or not are approximately equally likely. But if the glue falls into a SLD which is other than the SLD in investigation and shares the TLD with it, it is more likely to be signed. So a domain's operator should be cautious when setting its cross-domain glues. Fig. 12 also illustrates nearly half of domains are glue forgeable, which is largely due to their unsigned glues.

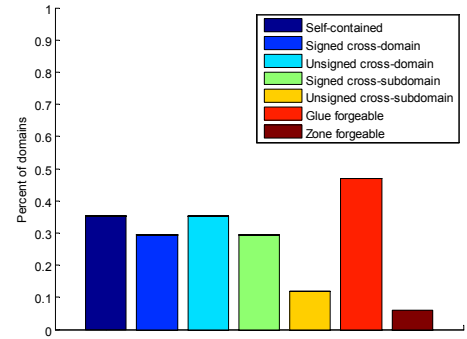

Fig. 12. The bailiwick location of glue and the status of DNSSEC deployment, and the glue and zone vulnerabilities.

To further investigate the glue and zone vulnerabilities, we show the DNSSEC and security status of glues in Fig. 13. We can see that a domain with all its glues signed is no guarantee of glue security. Instead, its glue is still forgeable if at least one of the glue's subdomain is subject to the NSEC3 Opt-Out vulnerability (as analyzed in Section III). A domain with at least one of its glues unsigned likewise does not necessarily means zone forgeability. If and only if the domain with forgeable glues uses NSEC3 and sets Opt-Out, it is vulnerable to zone forgeability.

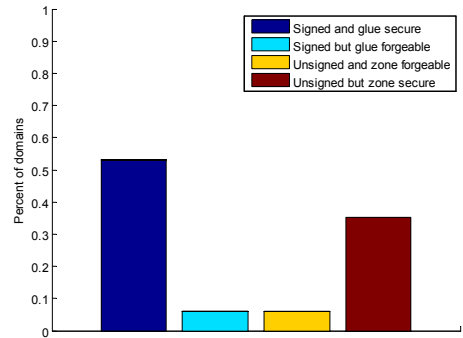

Fig. 13. The status of DNSSEC deployment and the glue and zone vulnerabilities.

## V. Related Work

DNS availability and interdependency issues are studied in [23] and [25]. Deccio et al. [23] proposed a formal model for analyzing the name dependencies inherent in DNS, and derived metrics to quantify the extent to which domain names affect other domain names. Yuan et al. [25] proposed metrics quantifying the Quality of Domain Name Service, and presented an analytical model of DNS proxy operations that offers insights into the design trade-offs of DNS infrastructure and the selection of critical DNS parameters. The threats of DNS, in the view of hackers and malware criminals, are surveyed in [31]. Shue et al. [28] examined DNS resolver behavior and usage, from query patterns and reactions to nonstandard responses to passive association techniques to pair resolvers with their client hosts.

The DNSSEC deployment challenges and impacts are extensive examined in recent years. Herzberg et al. [26] outlined the significant challenges and pitfalls that have resulted in severely limited DNSSEC deployment, and suggested directions for improvement and further research. Osterweil et al. [24] developed SecSpider, a DNSSEC monitoring system that helps identify operational errors in the DNSSEC deployment and discover unforeseen obstacles. Herzberg et al. [27] discussed the state of DNSSEC deployment and obstacles to adoption, and presented an overview of challenges and potential pitfalls of DNSSEC, including incremental deployment, long DNSSEC responses, trust model of DNS. Osterweil et al. [29] proposed to achieve robust DNSSEC verification with a new theoretical model, which treats operational deployments as Communities of Trust (CoTs) and makes them the verification substrate. Yang et al. [30] provided the systematic examination of the design, deployment, and operational challenges encountered by DNSSEC over the years, and revealed a fundamental gap between cryptographic designs and operational Internet systems.

## VI. Conclusions

This paper gives an extensive study of glue, a prevalent DNS configurations. Our analysis reveals the minimum glue records in terms of availability. And we also discover the security vulnerabilities of glue including three types of NSEC Opt-Out vulnerabilities. The measurements show the wide occurrences of glue redundancies and glue vulnerabilities in DNS practices.